# υ-SVR Polynomial Kernel for Predicting the Defect Density in New Software Projects

Cuauhtémoc López-Martín

*Department of Information Systems*
*Universidad de Guadalajara*
*México*
cuauhtemoc@cucea.udg.mx

Mohammad Azzeh

*Department of Software Engineering*
*Applied Science Private University*
*Amman, Jordan*
m.y.azzeh@asu.edu.jo

Ali Bou Nassif

*Department of Electrical and Computer Engineering*
*University of Sharjah*
*Sharjah, UAE*
anassif@sharjah.ac.ae

Shadi Banitaan

*Department of Mathematics, Computer Science and Software Engineering*
*University of Detroit Mercy*
*Detroit, MI, USA*
banitash@udmercy.edu

*Abstract*—An important product measure to determine the effectiveness of software processes is the defect density (DD). In this study, we propose the application of support vector regression (SVR) to predict the DD of new software projects obtained from the International Software Benchmarking Standards Group (ISBSG) Release 2018 data set. Two types of SVR (i.e., *ε-SVR* and *υ-SVR*) were applied to train and test these projects. Each SVR used four types of kernels. The prediction accuracy of each SVR was compared to that of a statistical regression (i.e., a simple linear regression, SLR). Statistical significance test showed that *υ-SVR* with polynomial kernel was better than that of SLR when new software projects were developed on mainframes and coded in programming languages of third generation.

## I. INTRODUCTION

An important product measure to determine the effectiveness of software processes is the defect density (DD), which has been defined as the total number of defects divided by the size of the software [2]. Software size has been mainly measured in either source lines of code or function points (FP) [3]. A defect is defined as an "imperfection or deficiency in a work product where that work product does not meet its requirements or specifications and needs to be either repaired or replaced.", and it is caused by a person committing an error [1]. A defect is associated to other SE knowledge area termed Software Quality [1].

On the other hand, the Software Engineering Management knowledge area is the application of management activities such as planning, and software project planning (SPP) addresses the activities undertaken to prepare for a successful software engineering project from the management perspective [1]. Software prediction (SP, also termed software *estimation*) belongs to SPP. SP has been used to predict variables related to software projects such as size, effort, duration or defects [4]. The types of software projects can be classified as new or maintained [5].

Several machine learning techniques have been applied in the SP field such as cased-based reasoning, neural networks, support vector regression (SVR), genetic programming, genetic algorithms, decision trees, and association rules [6].

A SVR is a type of Support Vector Machine (SVM) for regression tasks [7]. A SVR is useful to learn from non-linear relationships [8] [9], where non-linear relationships are common between software metrics and software defects [10]. SVR has been applied to predict the effort (person-months) [7], and duration (months) of software projects [21].

Software defect prediction models have been used to (1) identify software metrics, modules, classes, files or subsystems that are more likely to be defective (defect-prone) [12], (2) predict the number of defects [13], and predict DD [10] [11] [14] [15] [16] [17] [18] [19] [20].

As for DD prediction, the techniques applied have been neural networks [10] [17], fuzzy logic [10] [15], decision trees [17] [18], and statistical regressions [11] [14] [16] [19] [20].

In accordance with SVR specifically applied to defect prediction, we only identify three studies applying a SVR to defect-prone [12] [22] [23] rather than to DD.

The contribution of our study is the application of two types of SVR named *ε-SVR* and *υ-SVR* to predict the DD in new software projects using FP as the independent variable. Four types of kernels were used by type of SVR (i.e., linear, polynomial, radial basis function, and sigmoid).

The *ε-SVR* and *υ-SVR* were trained and tested with new software projects obtained from the International Software Benchmarking Standards Group (ISBSG) Release 2018.

The prediction accuracy obtained by SVR is compared to that of a statistical regression (i.e., simple linear regression, SLR) model because of any new proposed prediction model should at least outperform a statistical model [24].

The hypothesis to be investigated in our research is the following:

$H_1$: The DD prediction accuracy with the SVR is statistically better than the accuracy obtained by SLR when function points is used as the independent variable.

Section II presents studies where DD has been predicted. Section III describes the two types of SVR used to predict DD. Section IV describes the criteria followed to select the new software projects, the criteria to evaluate the prediction accuracy of prediction models, and the validation method to train and test the models. Section V shows the results of the prediction accuracies obtained from the models, whereas Section VI presents the conclusions, limitations, and future work.

## II. RELATED WORK

The following studies related to DD prediction were identified. They are chronologically described:

Verma and Kumar [14] use simple and multiple linear regression models to predict the DD of 62 open source software projects. They conclude that there is statistically significant level of acceptance for DD prediction using few repository metrics individually and jointly.

Yadav and Yadav [15] apply a fuzzy logic model for predicting DD at each phase of development life cycle of 20 software projects from the top most reliability relevant metrics of each phase. They conclude that the predicted DD are found very near to the actual defects detected during testing.

Mandhan et al. [16] predict DD by using simple and multiple regression models generated from seven different software static metrics (i.e., coupling, depth, cohesion, response, weighted methods, comments, and lines of code). They conclude that there is a significant level of acceptance for DD prediction with these static metrics individually and jointly.

Kumar [10] applies fuzzy logic and neural network to predict the DD of subsequent product releases of two software projects. He concludes that the neural network provides better results than a Fuzzy Inference System.

Rahmani and Khazanchi [11] apply simple and multiple regression models to predict DD of 44 open source software projects. They conclude that there is a statistically significant relationship between DD and number of developers and software size jointly; and that there is not significant relationship between DD and number of downloads.

Kutlubay et al. [17] apply neural networks and decision trees to predict DD. Data were obtained from nine projects. They conclude that decision trees outperform neural networks models.

Knab et al. [18] use a decision tree model to predict DD of seven releases of an open source web browser project. They conclude that (1) it is feasible to predict DD with acceptable accuracies with metrics from the same release, (2) to use lines of code has little predictive power with regard to DD, (3) size metrics such as number of functions are of little value for predicting DD, (4) it is feasible predict DD with satisfactory accuracy by using evolution data such as the number of modification reports, and that (5) change couplings are of little value for the prediction of DD.

Nagappan and Ball [19] apply statistical regression models to DD prediction using a set of relative code churn measures that relate the amount of churn to other variables such as component size and the temporal extent of churn. They conclude that absolute measures of code churn are poor predictors of DD, and that relative measures of code churn are highly predictive of DD.

Sherriff et al. [20] uses a multiple regression prediction model with 14 random in-process snapshots of a system. This model is then used to predict the DD of other six snapshots. This model is part of a suite of in-process metrics that leverages the software testing effort. They conclude that the resulting DD prediction is indicative of the actual system DD. Thus, it allows developers to take corrective actions earlier in the development process.

In our study, we propose the application of two types of SVR to predict the DD in new software projects using FP as the independent variable.

## III. SUPPORT VECTOR REGRESSION

SVM has its roots in statistical learning (Vapnik-Chervonenkis theory). In SVM, the optimization problem implies to find the maximum margin separating a hyperplane by correctly classify the training points as possible. An SVM model represents this optimal hyperplane with support vectors. An SVR, a type of SVM that can be applicable to regression problems, is characterized by the use of kernels, sparse solution, and Vapnik-Chervonenkis control of the margin and the number of support vectors [25].

An SVM has a set of slack variables $\mathcal{E}_i$, which correspond to the distance of the input vector $x^i$ from the decision hyperplane. These slack variables are associated with a parameter $C$, where C>0, which is used to control the over-training [7].

SVR is trained based on a symmetrical loss function (i.e., Vapnik's $\varepsilon$-insensitive), which equally penalizes high and low misestimates. This loss function corresponds to $\varepsilon$-insensitive of Vapnik, which a flexible tube of minimal radius is formed symmetrically around the estimated function, such that the absolute values of errors less than a certain threshold $\varepsilon$ are ignored both above and below the estimate. It permits that points outside the tube are penalized, but those points within the tube, either above or below the function, receive no penalty [25].

An SVR finds a function $f(x)$ that has most $\varepsilon$ deviation from the actually obtained target $y^i$ for the training data $x^i$, and at the same time is as flat as possible [26].

We apply two types of SVR named $\varepsilon$-*support vector regression* ($\varepsilon$-*SVR*) and $v$-*support vector regression* ($v$-*SVR*) [27] [28] [29] to predict DD of software projects.

The $\varepsilon$-*SVR* bases its prediction on the mentioned $\varepsilon$-*insensitive loss function*, whereas the $v$-*SVR* automatically minimizes the $\varepsilon$-*insensitive loss function*, and causes that the SVR formulation changes by using a new $v$ parameter whose value is between the [0,1] interval. The $v$ parameter is also used to control the number of support vectors, it implies that the $v$-*SVR* allows data compression and generalizes the prediction error bounds. In summary, there are two parameters $\varepsilon$ and C in $\varepsilon$-*SVR*, and $v$-*SVR* has the two parameters $v$ and $C$. Their relationship between these two types of kernels based on their parameters has been described as follows [30]: "the decision functions obtained by two methods are identical if the values of parameter $C$ are same, and the parameter $\varepsilon$ has a relationship with the parameter $v$".

The $\varepsilon$-*SVR* and $v$-*SVR* kernel functions used in this study are the following [21]:
- Linear: $K(x,z) = x \cdot y$, where $x$ and $z$ are data patterns
- Polynomial: $K(x,z) = (\gamma(x \cdot z) + c_0)^d$, where $\gamma$: slope parameter, $c_0$: trade-off between major terms and minor terms of the generated polynoms, $d$: polynom degree, and $x,z$: data patterns
- Radial basis function (RBF): $K(x,z) = \exp(-\gamma|x-z|^2)$, where $\gamma$ controls the radial basis function spread, and $x,z$: data patterns.

- Sigmoid: $K(x,z) = \tanh(\gamma(x \cdot z) + c_0)$, where $\gamma$ controls the radial base function spread, $c_0$: independent term, and $x,z$: data patterns.

## IV. METHODOLOGY

In this section, we describe the source of projects as well as the followed criteria to select them from an international repository. Moreover, the criteria to evaluate the prediction accuracy of prediction models are described. Finally, the validation method used to train and test the models is justified and applied.

### A. Data Set of new software projects

The new software projects used in our study were obtained from the public ISBSG data set Release 2018. This release contains 8,261 projects developed between the years 1989 and 2016. The data of these projects were submitted to the ISBSG from 26 different countries [31].

Table I describes the criteria applied for selecting the new projects according to attributes suggested in the ISBSG guidelines [5].

TABLE I. CRITERIA FOR SELECTING NEW SOFTWARE PROJECTS FROM THE ISBSG DATA SET (A AND B MEAN HIGHER QUALITY)

| Attribute | Selected value(s) | Number of projects |
|---|---|---|
| Data quality rating | A and B | 7,780 |
| Unadjusted function point rating | A and B | 6,429 |
| Defect density | Not null | 669 |
| Type of development | New | 211 |
| Development platform | Not null | 170 |
| Language type | Not null | 167 |
| Functional sizing methods | IFPUG V4+ and NESMA | 140 |

Two of the final 140 new software projects of Table I were excluded because they had a defect density value lower than one (zero and 0.1 values). In Table II, the 138 new projects are classified by develop platform and language type.

TABLE II. NEW PROJECTS CLASSIFIED BY DEVELOPMENT PLATFORM (MF: MAINFRAME, MR: MIDRANGE, MULTIPLATFORM, PC: PERSONAL COMPUTER) AND LANGUAGE TYPE

| Development platform | Language type | Number of projects |
|---|---|---|
| MF | 3GL | 24 |
| | 4GL | 8 |
| | ApG | 4 |
| MR | 3GL | 11 |
| | 4GL | 10 |
| Multi | 3GL | 31 |
| | 4GL | 13 |
| PC | 3GL | 11 |
| | 4GL | 26 |
| Total | | 138 |

The independent variable used in our study is FP, which is a composite value calculated from two data functions (internal logical file, and external interface files), and three transactional functions (external inputs, external outputs, external inquiries) [4], whereas in the ISBSG the DD is defined as defects by 1000 FP and calculated as follows [32]:

*Total defects delivered * 1000 / functional size*

DD is defined as the number of defects by 1000 functional size units of delivered software in the first month of use of the software. It is expressed as *defects by 1000 function points*.

In our study, with the goal of obtaining a better generalization from our conclusions based on statistical significance, the three datasets from Table II higher than twenty projects were selected to be analyzed (i.e., 24, 31 and 26 projects). A scatter plot (FP vs. DD) was generated by data set. The three scatter plots showed skewness and heteroscedasticity, and outliers. Table III includes four statistical distribution tests commonly used to data normality (the Chi-squared test was not possible apply it to two datasets because it needs at least thirty data) applied for dependent and independent variables by data set. In accordance with p-values of Table III, we can reject the idea that variable comes from a normal distribution with 99% confidence.

TABLE III. NORMAL STATISTICAL TESTS BY DATA SET (DP: DEVELOPMENT PLATFORM, LT: LANGUAGE TYPE, NP: NUMBER OF PROJECTS, FP: FUNCTION POINTS, DD: DEFECT DENSITY)

| DP | LT | NP | Variable | Statistical test | | | |
|---|---|---|---|---|---|---|---|
| | | | | Chi-squared | Shapiro-Wilk | Skewness | Kurtosis |
| MF | 3GL | 24 | FP | --- | 0.0006 | 0.0223 | 0.0132 |
| | | | DD | --- | 0.0017 | 0.0461 | 0.0606 |
| Multi | 3GL | 31 | FP | 0.0000 | 0.0000 | 0.0264 | 0.2732 |
| | | | DD | 0.0000 | 0.0000 | 0.0000 | 0.0000 |
| PC | 4GL | 26 | FP | --- | 0.0000 | 0.0018 | 0.0005 |
| | | | DD | --- | 0.0000 | 0.0002 | 0.0000 |

Thus, data of the three datasets were normalized by applying the natural logarithm (*ln*) as suggested by Kitchenham [24]. Afterwards, outliers were identified by means of studentized residuals greater than 2 in absolute value (studentized residuals measure how many standard deviations each observed value of DD deviates from the SLR model fitted using all of the data except that observation). Row data set of Table II consisting of 24 projects was normalized. This data set had three outliers, which were excluded. Figure 1 depicts the final data set consisting of 21 new software projects.

Data of Figure 1 had a coefficient of correlation equal than -0.80 and a p-value equal than 0.0000, indicating statistically significant non-zero correlation at the 99% confidence level. Figure 1 shows that the higher the value of *FP*, the lower the *defect density* is in the first month of use of the software. This pattern coincides with that reported in some studies ([2] [33] [34]).

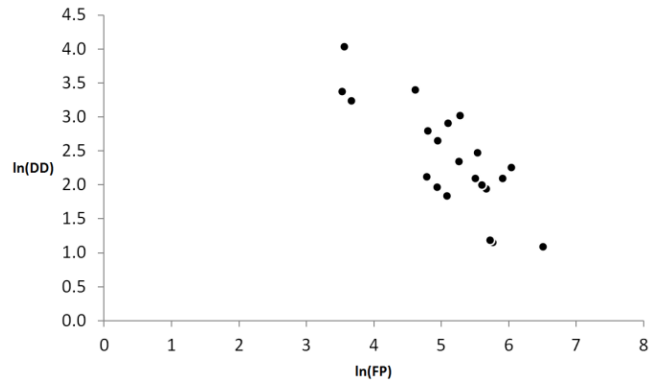

Figure 1. Normalized data set without outliers

In accordance with the other two data sets consisting of 31 and 26 projects, they did not presented statistically significant non-zero correlation when all outliers were removed. Therefore, in our study only the following data set was used:
- Twenty-four new software projects developed in mainframes and coded in programming languages of third generation.

SLR Equation 1 was that one generated from this data set. Its ANOVA p-value was equal to 0.0000, that is, there is a statistically significant relationship between the variables at the 99% confidence level, whereas its coefficient of determination ($r^2$) was equal to 0.64, which means that the SLR as fitted explains 64% of the variability in the DD.

$$\ln(DD) = 6.42311 - 0.787475*\ln(FP) \quad (1)$$

### B. Accuracy Criteria

The magnitude of relative error (MRE) is the accuracy measure which is the mostly used to evaluate software prediction models based on machine learning [6], the magnitude of error relative to the estimate (MER) has been recommended over the MRE [35], and absolute residuals (AR) has also been recommended over MRE since MRE is a non-asymmetric measure [36]. In our study, we used the three mentioned accuracy measures such that our results can be compared with future studies.

The equations for MRE, MER and AR are presented in (2) (3) and (4), respectively.

$$MRE_i = \frac{|Actual\ DefectDensity_i - Predicted\ DefectDensity_i|}{Actual\ DefectDensity_i} \quad (2)$$

$$MER_i = \frac{|Actual\ DefectDensity_i - Predicted\ DefectDensity_i|}{Predicted\ DefectDensity_i} \quad (3)$$

$$AR_i = |Actual\ DefectDensity_i - Predicted\ DefectDensity_i| \quad (4)$$

The MRE, MER and AR are calculated for each observation $i$, the DD of which are predicted. The aggregation of the MRE, MER, and AR over multiple observations (N) can be obtained by their mean (MMRE, MMER and MAR) as shown in (5) (6), and (7).

$$MMRE = (1/N)\sum_{i=1}^{N} MRE_i \quad (5)$$

$$MMER = (1/N)\sum_{i=1}^{N} MER_i \quad (6)$$

$$MAR = (1/N)\sum_{i=1}^{N} AR_i \quad (7)$$

The MdMRE, MdMER, and MdAR correspond to the median of the MREs, MERs and ARs, respectively. The accuracy of a prediction model is inversely proportional to any of the accuracy measures mentioned above.

### C. Training and Testing the Models

Since leave-one-out cross-validation (LOOCV) has been recommended as validation method because it removes the random selection for train and test sets by avoiding their nondeterministic selection [4], we used it to train and test the two types of SVR and the SLR. That is, since a data set with 21 new software projects was used, then 21 SLR were generated when a LOOCV was performed.

In accordance with ε-SVR training, the dual optimization problem formulated in Equation 8 was solved (considering that, under the parameters $C>0$ and $\varepsilon>0$, the $\varepsilon$-SVR is a regression model that solves the optimization problem) [8]:

$$\min_{w,b,\xi,\xi^*} \frac{1}{2} w^T w + C \sum_{i=1}^{l} \xi_i + C \sum_{i=1}^{l} \xi_i^* \quad (8)$$

Subject to $w^T \phi(x^i) + b - y^i \leq \epsilon + \xi_i,$
$y^i - w^T \phi(x^i) - b \leq \epsilon + \xi_i^*,$
$\xi_i, \xi_i^* \geq 0, i = 1,..,l$

Afterwards, the decision function described in (8) was obtained. Equation 9 corresponds to the approximated decision function once the dual form of the optimization problem from the method of Lagrange multipliers was solved [7]:

$$f(x) = \sum_{i=1}^{l}(-\alpha_i + \alpha_i^*)K(x^i, x) + b \quad (9)$$

In the testing phase for the classification, a test pattern (i.e., a new software project) was used for the decision function obtained from the ε-SVR optimal solution.

The $\varepsilon$ parameter is replaced by the $v$ parameter in the procedure to train and test the data set of our study when applied the $v$-SVR. The rest of the procedure by using the $v$-SVR was the same than that performed for ε-SVR [30].

## V. RESULTS

A suitable statistical test for comparing the accuracy results showed in Table IV is selected based on the number of data sets to be compared, data dependence, and data distribution [37]: two data sets at the same time will be compared by accuracy measure (one from SLR and one from each SVR kernel), data are dependent because DD by project was predicted applying the two types of models (SLR and each SVR), whereas for data distribution, normality test by a new data set obtained from the differences between prediction accuracy measures obtained between the two models, are performed.

Table V includes the p-values obtained by comparing SLR accuracy with that of each kernel whose accuracy measure has been better (lower) than that of the SLR. The statistical tests used were *t-paired* (a parametric statistical test) for MMRE, MMER and MAR, and Wilcoxon (a nonparametric statistical test) for MdMRE, MdMER and MdAR.

Based on Tables IV and V, it can be observed that the $v$-SVR polynomial kernel was the unique kernel which resulted statistically better than SRL.

TABLE IV. SLR AND SVR MODELS ACCURACY RESULTS (L: LINEAR, P: POLINOMIAL, RBF: RADIAL BASIS FUNCTION, S: SIGMOIDAL)

| Model | Kernel | MMRE | MdMRE | MMER | MdMER | MAR | MdAR |
|---|---|---|---|---|---|---|---|
| SLR | --- | 0.20 | 0.18 | 0.19 | 0.21 | 0.42 | 0.49 |
| $\varepsilon$-SVR | L | 0.21 | 0.16 | 0.18 | 0.19 | 0.41 | 0.44 |
|  | P | 0.19 | 0.13 | 0.16 | 0.15 | 0.38 | 0.32 |
|  | RBF | 0.27 | 0.17 | 0.20 | 0.20 | 0.48 | 0.49 |
|  | S | 0.31 | 0.21 | 0.29 | 0.20 | 0.65 | 0.46 |
| $v$-SVR | L | 0.21 | 0.19 | 0.19 | 0.22 | 0.44 | 0.47 |
|  | P | 0.17 | 0.14 | 0.16 | 0.17 | 0.36 | 0.40 |
|  | RBF | 0.23 | 0.14 | 0.20 | 0.16 | 0.45 | 0.34 |
|  | S | 0.33 | 0.23 | 0.26 | 0.20 | 0.65 | 0.50 |

TABLE V. STATISTICAL TEST RESULTS (P-VALUES) BY COMPARING THE SRL ACCURACY WITH ACCURACY OF EACH KERNEL

| Kernel | MMRE | MdMRE | MMER | MdMER | MAR | MdAR |
|---|---|---|---|---|---|---|
| $\varepsilon$-SVR, L | --- | 0.9999 | 0.4683 | 0.9652 | 0.7370 | 0.9839 |
| $\varepsilon$-SVR, P | 0.2412 | 0.2968 | 0.0621 | 0.0700 | 0.1386 | 0.1539 |
| $\varepsilon$-SVR, RBF | --- | 0.2064 | --- | --- | --- | --- |
| $\varepsilon$-SVR, S | --- | --- | --- | 0.0515 | --- | --- |
| **$v$-SVR, P** | 0.0254 | 0.0151 | 0.0194 | 0.0136 | 0.0117 | 0.0094 |
| $v$-SVR, RBF | --- | 0.9169 | --- | 0.8224 | --- | 0.8665 |

Since the smallest p-value amongst the statistical distribution tests performed for the three data sets is less than 0.01 in Table VI, we can reject the idea that each data set of differences comes from a normal distribution with 99% confidence. It means that the comparison between SLR accuracy and $v$-SVR polynomial kernel should be based on their medians (i.e., based on Wilcoxon test).

TABLE VI. STATISTICAL TEST RESULTS (P-VALUES) BY NORMALITY DISTRIBUTION TEST

| Normality test | p-values | | |
|---|---|---|---|
|  | MRE | MER | AR |
| Shapiro-Wilk | 0.0006 | 0.0032 | 0.0348 |
| Skewness | 0.0125 | 0.0328 | 0.1130 |
| Kurtosis | 0.0011 | 0.0116 | 0.1799 |

Table VII shows the final parameters for $v$-SVR polynomial kernel which generated those MdMRE = 0.14, MdMER = 0.17, and a MdAR = 0.40 showed in Table IV. The Coefficient coef0 equal to zero means that the polynomial kernel is homogeneous, whereas the shrinking heuristic saves training time by identifying and removing some bounded elements (like $C$ or $\alpha_i = 0$), which leads to a smaller optimization problem [21].

TABLE VII. PARAMETERS FOR U-SVR POLYNOMIAL KERNEL

| Parameter | Value |
|---|---|
| Kernel Type | $(\gamma x^i x + coef0)^{degree}$ |
| $\gamma$ | 0.2 |
| coef0 | 0 (zero means the kernel is homogeneous) |
| Degree | 2 |
| $C$ | 1.0 |
| $v$ | 0.09 |
| Shrinking heuristic | Yes |

## VI. CONCLUSIONS, LIMITATIONS AND FUTURE WORK

DD is a software quality attribute that gives the reliability measure of the software product [10]. Therefore, we proposed the application of two types of support vector regression (SVR) termed ε-*SVR* and υ-*SVR* to predict the DD of new software projects. Each type of SVR used linear, polynomial, radial basis function, and sigmoid kernels.

The prediction accuracy of each type of SVR was compared to that of a statistical regression. This comparison was based on the MRE, MER and AR, which are three commonly accuracy measures used in software prediction field.

Based on statistical tests, the following hypothesis derived from that formulated in the Introduction section of this study, can be accepted:

$H_1$: The accuracy of the defect density of new software projects with a $v$-*SVR* with polynomial kernel is statistically better at the 95% confidence level, than the accuracy obtained by SLR when function points is used as the independent variable.

We can conclude that a $v$-*SVR* with polynomial kernel could be used for predicting the DD of new software projects developed in mainframes and coded in programming languages of third generation

In comparing our study with previous ones:

a) We did not find any study applying SVR to predict DD from projects obtained from the ISBSG data set.

b) Researchers have approached their efforts in analyzing how DD changed with size [2] [33] [34]. Figure 1 of our study coincides with conclusions of previous studies in the sense that large projects exhibited lower DD than medium and small projects. This pattern may be explained because larger projects tend to be developed more carefully [34]. A recent study published in 2018, reports that larger modules tend to have more defects but have a lower DD [38].

Regarding limitations of our study, although the last version of the ISBSG release 2018 consists of 2,557 new software projects of the total (8,261 projects), after we followed the criteria suggested by the ISBSG for selecting the data sets for new software projects, we could only use a data set of 21 projects to train and test the models.

A validation threat of our study is related to the independent variable used (i.e., FP). FP is also predicted, thus, the accuracy of the $v$-*SVR* with polynomial kernel also depends of the size estimation accuracy.

Future work will be related to the application of support vector regression for predicting the DD in software maintenance projects.

ACKNOWLEDGEMENT

The authors would like to thank the CUCEA, Universidad de Guadalajara, and the Consejo Nacional de Ciencia y Tecnología (CONACyT), México; the Applied Science Private University, Amman, Jordan;  University of Sharjah, Sharjah, UAE; and the University of Detroit Mercy, Michigan, USA, for their support during the development of this work.